# Synchrotron radiation studies of spectral response features caused by Te inclusions in a large volume coplanar grid CdZnTe detector

C.C.T Hansson[a], Alan Owens[b], F. Quarati[c], A. Kozorezov[d], V. Gostilo[e], D. Lumb[f].

(a) Solar System Missions Division. ESA/ESTEC, Noordwijk, Netherlands
(b) Office for Support to New Member States, ESA/ESTEC, Noordwijk, Netherlands
(c) Faculty of Applied Sciences, Technical University of Delft, Delft, Netherlands
(d) Department of Physics, University of Lancaster, Lancaster, England, UK
(e) Baltic Scientific Instruments, 26 Ganibu dambis, PO Box 33, Riga LV-1005, Latvia
(f) Astrophysics & Fundamental Physics Mission Division, ESA/ESTEC, Noordwijk, Netherlands

**Abstract**

We report preliminary results from a synchrotron radiation study of Te inclusions in a large volume single crystal CdZnTe (CZT) coplanar-grid detector. The experiment was carried out by probing individual inclusions with highly collimated monochromatic X-and γ-ray beams. It was found that for shallow X-ray interaction depths, the effect of an inclusion on measured energy loss spectra is to introduce a ~10% shift in the peak centroid energy towards lower channel numbers. The total efficiency is however not affected, showing that the net result of inclusions is a reduction in the Charge Collection Efficiency (CCE). For deeper interaction depths, the energy-loss spectra shows the emergence of two distinct peaks, both downshifted in channel number. We note that the observed spectral behavior shows strong similarities with that reported in semiconductors which exhibit polarization effects, suggesting that the underlying mechanism is common.

## 1. Introduction

CdZnTe (CZT), in view of its large average atomic number (~49) and wide band gap (1.572eV), has long been recognised for its potential as a room temperature X-ray and γ-ray detection medium [1], potentially having a spectroscopic performance close to that of Si. Until recently the spectroscopic performance of CZT detectors have been impaired by charge trapping due to crystal defects such as voids, twins, grain boundaries, and inclusions [2]. Recent improvements in crystal growth have largely removed macroscopic defects - the spectroscopic performance now being limited by the existence of the Te inclusions and precipitates. Both inclusions and precipitates are regions of Te incorporated into the CZT crystal structure. The formation of precipitates has been attributed to the cooling process and originates from the retrograde solubility in CZT [3], while the creation of Te inclusions has been attributed to morphological instabilities at the growth interface leading to the capture of Te rich droplets from the melt [3]. Typical diameters of Te precipitates range from (10-20) nm while inclusions have a size in the range 1-50μm [3], making the effects of inclusions more significant. The mechanism(s) leading to reduced detector response due to inclusions are however currently not well understood [4-6]. Consequently, considerable effort is now being expended to correlate the size and density distributions of Te inclusions with spectroscopic performance [4].

## 2. Detector and experimental setup

The detector was fabricated from a 15×15×10 mm$^3$ single CZT HPB grown crystal manufactured by Yinnel. Au electrodes were patterned onto the crystal faces in a coplanar arrangement on the top face and a single planar cathode on the opposite face (Fig. 1-left). All connections to the electrodes were achieved by wire bonding. The detector was mounted on a dielectric substrate which, in turn, is housed in an Al housing (Fig. 1 – right).

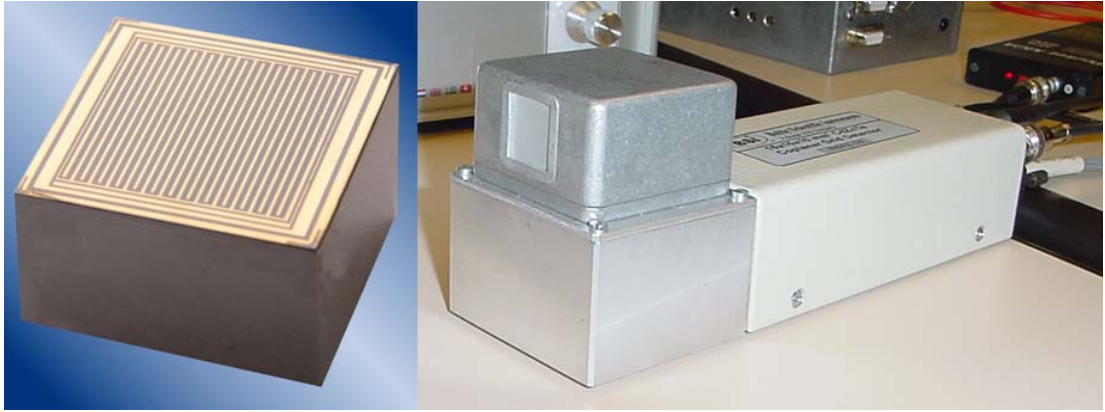

Fig. 1. Left - Photograph of the coplanar grid detector. Right – Photograph of the fully detector assembly.

For the detector readout, both grids are connected to a standard resistive feedback preamplifiers. Separate charge sensitive amplifers are connected to the collecting and non-collecting grids, and after subtraction of the latter the remainder of the signal processing chain comprises a conventional Ortec 671 spectroscopy amplifier. The electrical characterisation of the device has been described elsewhere [7]. A bias of 1900V was applied, modulated with an intergrid bias of 80V. The detector was irradiated through a thin Be window onto the anode side (coplanar grid). X- and γ-ray characterization of the device was carried out at the Hamburger Synchrotronstrahlungslabor (HASYLAB) radiation facility in Hamburg, Germany and the European Synchrotron Research Facility (ESRF) in Grenoble, France. The hard X-ray response was measured on beam line X1 at HASYLAB. This station uses a double crystal monochromator to produce X-rays in the range 10-100keV, with an approximate energy resolution of 1eV at 10keV rising to 20eV at 100keV. For the measurements carried out here a beam size of (10×10) μm was used. The γ-ray measurements were carried out on beam line ID15 at the ESRF using a (50×50) μm beam. This beam line uses a double crystal monochromator to produce X-rays in the (30-1000) keV range. An asymmetrical multipole wiggler (AMPW) and a superconducting wavelength shifter (SCWS) were also used in order achieve usable fluxes over the allowable energy range. At both beamlines, the detector was located on a high precision X-Y table, as can be seen in Fig. 2, and the detector surface raster scanned using the incoming X-ray beam.

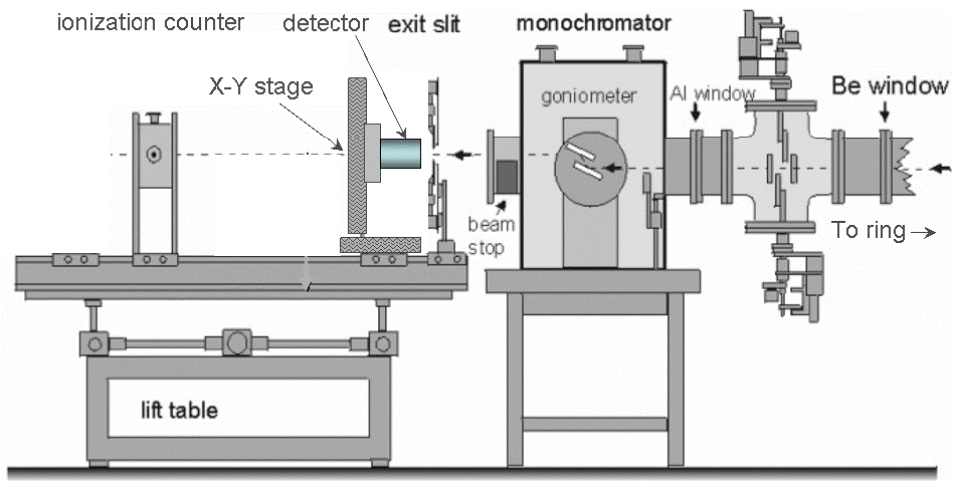

Fig. 2. Illustration of the HASYLAB experimental setup used to characterize the detector at hard X-ray wavelengths [8].

### 3. Experimental results

*3.1 Spectral response at 60 keV and size distribution of Te inclusions*

A coarse full area raster scan of the detector was conducted at 60 keV in order to indentify regions of interest. At each point in the scan, energy–loss spectra are acquired, which represent the collected charge at that position. The spectra were then quantified in terms of the peak channel of the photopeak, its width and total counts in the peak. Three regions, with a total surface area of 5 mm$^2$, were subsequently identified and finely scanned. At 60 keV the 1/e absorption depth in CZT is 0.33mm. The shallow interaction depth in combination with anode illumination and the chosen dynamic gain factor mean that the contribution of both electrons and holes drift in the vicinity of the inclusions may be important. A total of 21 inclusions, of various sizes within regions of interest were analysed, Data were visualised using spatial distribution of peak centroid position. As can be seen in Fig. 3 the spectral response, measured across the inclusions showed a peak shift to lower channel numbers in moving from edge to centre of the inclusions. The total event rate with position across the inclusion was not altered, showing that the presence of the inclusion does not affect the overall efficiency but does result in a reduction of the charge collection efficiency (CCE). By using the width of the peak centroid shift profile (see Fig. 3 – bottom) as an indicator of the inclusion size, the size-to-shift distribution was investigated for the 21 inclusions observed and can be seen in Fig. 4. Surprisingly, the inclusion range of sizes observed is much larger than is the 1-50µm expected. No direct correlation between the size and channel shift distribution was observed.

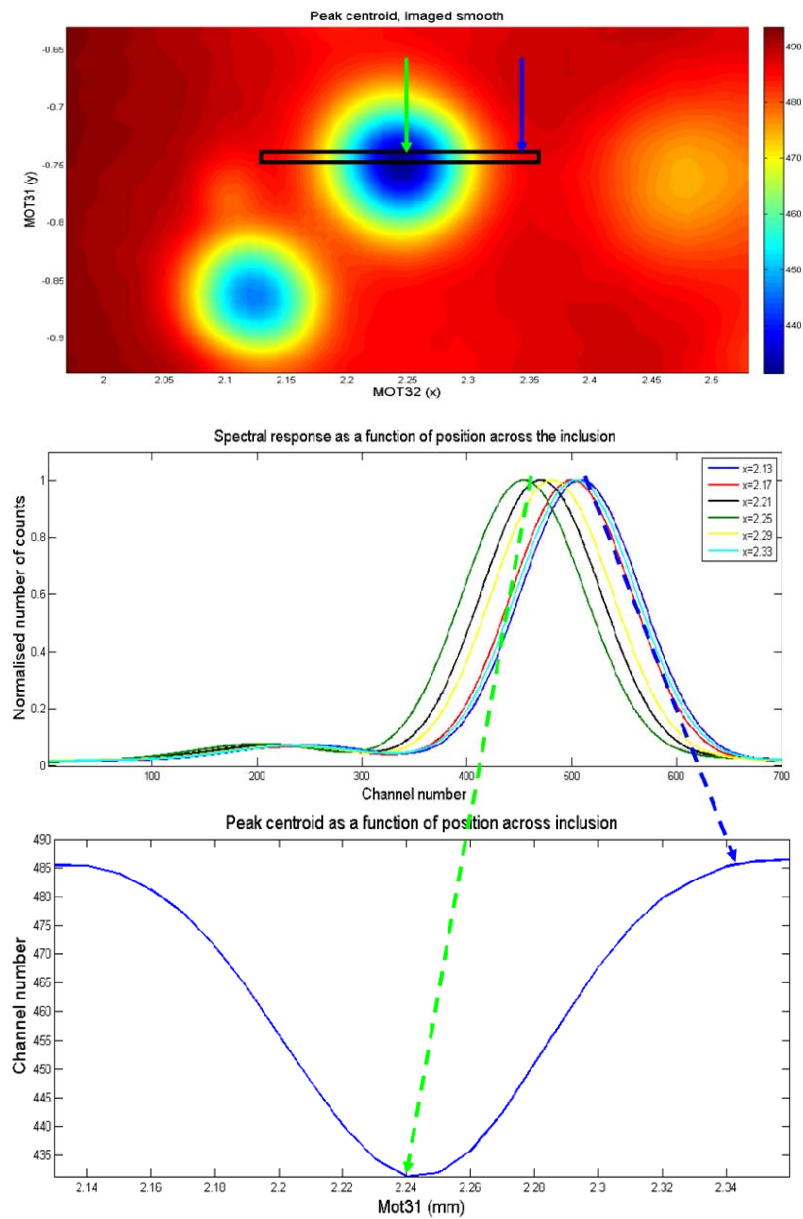

Fig. 3. Top – Image of one of the regions of interest scanned at 60 keV with three inclusions clearly visible. The colour scale correspond to peak channel number with red being high and blue low. Centre – The spectral response observed when moving across the inclusion (line in top image). Bottom – Peak channel number visualisation across the inclusion. The inclusion caused a shift of the peak channel number in the spectral response.

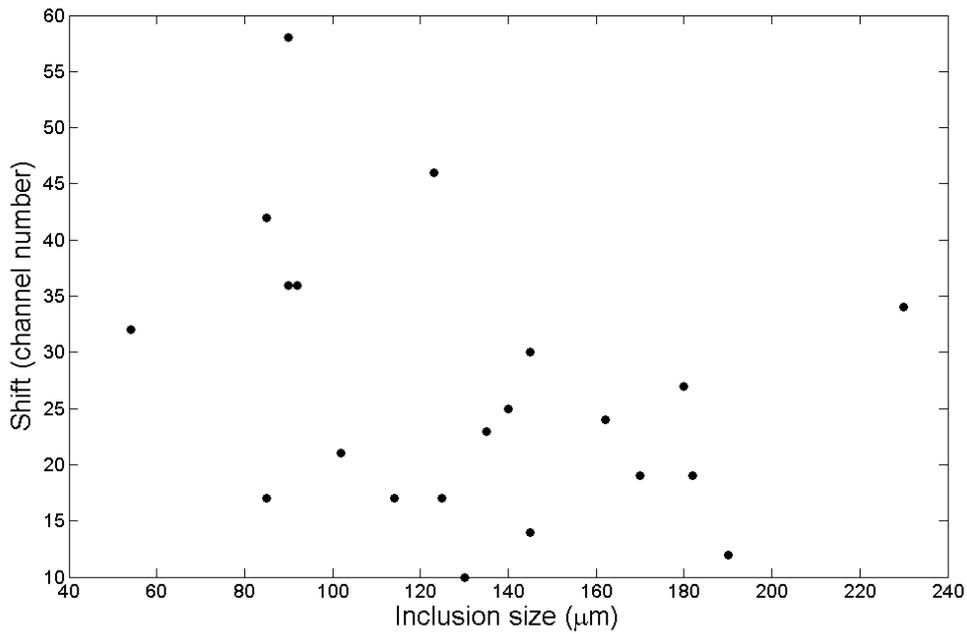

Fig. 4. Plot of the inclusion size as a function of shift in channel number. No direct correlation can be observed.

*3.2 Spectral response at 180 keV of Te inclusions*

The entire surface area of the detector was raster scanned at 180 keV and the spectral response across the inclusions observed. At 180 keV the 1/e absorption depth in CZT is 7.3mm. The much deeper interaction depth increases the probability that the inclusions observed have been traversed by the electron charge cloud. Any inclusion encountered by the electron cloud will result in increased electron trapping and lead to a reduced charge being induced on the electrode once the cloud reaches the near field region. A larger volume of the detector, and hence a larger number of inclusions, is therefore imaged at 180 keV than at 60 keV. The spectral response across an inclusion at this energy was observed and can be seen in Fig. 5.

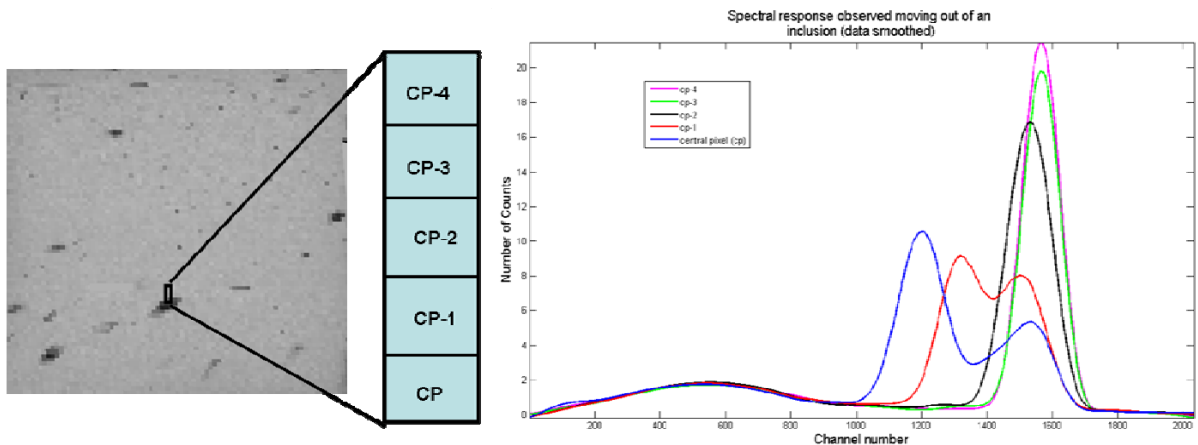

Fig. 5. Spectral response observed at 180 keV when scanning the beam across the inclusion. A significantly different spectral response is observed compared to the 60 keV scenario.

The spectral response showed a gradual shift in the peak centroid towards lower channel number when entering the inclusion leading to two distinct peaks both downshifted when the beam is located entirely inside the inclusion. The significantly different effect on the spectral response observed at 60 keV and 180 keV suggests that the net effect of the Te inclusion on the transport of the created charge carriers is dependent on which charge carrier type is traversing the inclusion. The emergence of dual peaks at 180 keV also suggests that the position of the photon interaction, with respect to the location of the inclusion, can have a significant impact on the charge measured for that event.

*3.3 Beam size variation*

The spectral response caused by an inclusion when using different beam sizes was investigated by centering the 60 keV beam on an inclusion and gradually increasing the beam size from (10x10)μm to

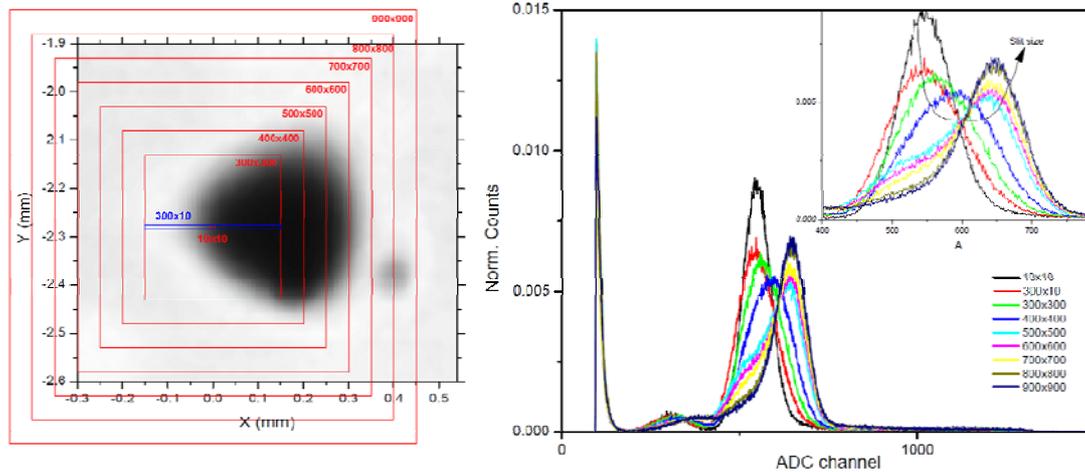

Fig. 6. The spectral response observed when increasing the beam size to inclusion size ratio. A gradual increase in peak channel number is observed with increased beam size.

(900x900) μm. The result can be seen in Fig. 6. When the beam cross section area is less than that of the inclusion and the beam position is such that the projected cross-section is inside the inclusion cross-section (10x10μm), the expected response, with a single peak shifted to lower channel number, was observed. As the beam size is increased, we see a broadening of the peak combined with a shift to higher channel number finally resulting in a peak positioned at a higher channel number but with a "low energy" tailing (900x900μm). This phenomenon can be explained as part of the beam being aligned with the inclusion and part of the beam hitting a non inclusion affected part of the detector and the spectral response being a convolution of the two. From the experiment the importance of size and density distribution of Te inclusion on the spectral response of the CZT crystal can be inferred. It also shows the importance of the beam size to inclusion size ratio when using this technique.

**4. Discussion**

The best quality CZT crystals currently available contain Te inclusions of size up to 10 μm with typical density of $\rho \sim 10^4$ cm$^{-3}$. For 10x10 μm$^2$ beam the probability of finding a single inclusion within the exposed volume in a 1 cm thick crystal is 0.01, becoming much less for shallower absorptions. At 60 keV, the probability of finding a single inclusion is reasonably high using a 900x900 μm$^2$ beam, but much smaller for two or more inclusions. This means that all features seen in our experiments can most likely be attributed to a single inclusion. Te inclusions are known to enhance the carrier trapping, however the exact mechanisms are not known. In most recent work by Bale [9], the heterogeneous material is represented by a homogenized CdZnTe crystal whose effective electron attenuation length incorporates the additional *uniform* electron trapping caused by the inclusions. Our results show that this approach can only be justified for a large diameter of the exposed area or for a full illumination of the detector where the effect of a large number of inclusions may be discussed in the context of an effective medium approximation. The fact that the experimentally determined size of inclusions significantly exceeds earlier reported data gathered from optical measurements indicates that Te inclusions cause a significant distortion of internal electric field deflecting electrons and holes from straight drift trajectories - meaning they are exposed to bulk traps rather than being trapped by inclusions themselves. Also, the absence of a significant low tailing in the signal shapes supports this conclusion. In other words, the numerous traps attached to the Te inclusions are already saturated.

Te is a semimetal. Te inclusions in a CZT matrix are neutral but highly polarizable in an external electric field, due to an induced dipole moment in the inclusion by that electric field. The electric field outside spherical Te inclusion in the bulk of coplanar grid detector (further away from near field region of the grid structure) can be represented as a superposition of the uniform external field and the field of an induced dipole moment. As a result the electric field around the inclusion will be distorted resulting in change of carrier drift trajectories. Similar effects occur in detectors showing the effect of polarization due to space charge build-up [10]. In the latter situation the internal electric field around the polarised volume is strongly distorted, and with electric field near the pinch point weakened a significant shift in peak channel position and even doubling of the line was observed and theoretically simulated. In the polarized detector, the experimentally determined extent of the "damaged" area exceeded the real size of the exposed spot indicating a long range distortion of internal field around the polarized volume. However, while the effects of polarization and inclusions are similar, the underlying mechanisms are different and work is ongoing to further develop the model and simulate line shapes and their transformation. Nevertheless, our results indicate that the major effect is related to carrier drift slowing down in the vicinity of Te inclusion leading to enhanced trapping on traps in the bulk

## 5. Conclusions

A CZT crystal, configured in a coplanar grid configuration, was raster scan illuminated using 60 keV and 180 keV X-ray beams. A significantly different spectral response was observed at the two X-ray energies, suggesting that the net effect of Te inclusions is dependent of the charge carrier type traversing the inclusion, as well as the depth of the inclusion. Initial modelling and simulations suggest this is due to the variations in the electrostatic potential caused by the inclusion. By varying the beam size the importance of the inclusion size and density distribution on the spectral response was inferred. For a fixed inclusion size, this effect will become more significant for pixelated devices where the relative ratio of non-affected and affected regions will be shifted, causing greater spectral distortions. An absence of correlation between the inclusions size and peak channel shift also suggests that not only the size and density distribution of inclusion will be of importance but also their respective depth inside the crystal. No correlation between the inclusion size and peak channel number shift could be found.

**Acknowledgements**

We thank John van der Biezen (ESA-ESTEC) and Adam Webb (HASYLAB) for their help and support leading up to and during the testing.

**References**

[1] D.S. McGregor and H. Hermon, "Room-temperature compound semiconductors radiation detectors", Nucl. Instr. And Meth. A 395, 101-124, 1997.
[2] A.E. Bolotnikov et al. "Extended defects in CdZnTe radiation detectors", IEEE Trans. Nucl. Sci. 56, 4, 1775-1783, 2009.
[3] C. Szeles et al. "Advances in the crystal growth of semi-insulating CdZnTe for radiation detector applications", Trans. Nucl. Sci. 49, 5, 2535-2540, 2002.
[4] A.E. Bolotnikov et al. "The effect of Te precipitates on characteristics of CdZnTe detectors", Proc. of SPIE, 6319, 631903, 2006.
[5] A.E. Bolotnikov et al. "Performance-limiting defects in CdZnTe detectors", IEEE Trans. Nucl. Sci. 54, 4, 821-827, 2007.
[6] A.E. Bolotnikov et al. "Te inclusions in CdZnTe detectors : new method for correcting their adverse effects", IEEE Trans. Nucl. Sci. 57, 2, 910-919, 2010.
[7] A. Owens et al. "Hard X- and γ-ray measurements with a large volume coplanar grid CdZnTe detector", Nucl. Instr. And Meth. A 563, 242-248, 2006.
[8] http://hasylab.desy.de/facilities/doris_iii/beamlines/x1_roemo_ii/index_eng.html
[9] D.Bale, J. Appl. Phys, **108**, 024504 (2010)
[10] A. Kozorezov, V. Gostilo, A. Owens, F. Quarati, M. Shorohov, M.A. Webb, J.K. Wigmore, J. App. Phys. 108, 064507-10 (2010).